\documentclass[journal=jacsat,manuscript=article]{achemso}

\usepackage[version=3]{mhchem} 
\usepackage{float}
\usepackage{graphicx}
\usepackage{amsmath}
\usepackage{amssymb}
\usepackage{physics}
\usepackage{ragged2e}
\usepackage{tabularx}
\usepackage{bm}
\usepackage{subfig}
\usepackage{booktabs}
\usepackage{arydshln}
\usepackage[]{algorithm}
\usepackage{algpseudocode}

\newcommand{\floor}[1]{\lfloor #1 \rfloor}
\newcommand{\ceil}[1]{\lceil #1 \rceil}
\newcommand\ddfrac[2]{\frac{\displaystyle #1}{\displaystyle #2}}
\DeclareMathOperator*{\argmin}{arg\,min}

\author{Teemu H\"ark\"onen}
\affiliation[LUT University]
{LUT School of Engineering Science, LUT University, Lappeenranta, Finland}
\email{teemu.harkonen@lut.fi}
\author{Lassi Roininen}
\affiliation[LUT University]
{LUT School of Engineering Science, LUT University, Lappeenranta, Finland}
\author{Matthew T. Moores}
\affiliation[University of Wollongong]
{National Institute for Applied Statistics Research Australia, University of Wollongong, Australia}
\author{Erik M. Vartiainen}
\affiliation[LUT University]
{LUT School of Engineering Science, LUT University, Lappeenranta, Finland}

\title[Bayesian quantification for CARS]
  {Bayesian quantification for coherent anti-Stokes Raman scattering spectroscopy}

\begin{document}

%



%

\begin{abstract}
  We propose a Bayesian statistical model for analyzing coherent anti-Stokes Raman scattering (CARS) spectra.
  Our quantitative analysis includes statistical estimation of constituent line-shape parameters, underlying Raman signal, error-corrected CARS spectrum, and the measured CARS spectrum.
  As such, this work enables extensive uncertainty quantification in the context of CARS spectroscopy.
  Furthermore, we present an unsupervised method for improving spectral resolution of Raman-like spectra requiring little to no \textit{a priori} information.
  Finally, the recently-proposed wavelet prism method for correcting the experimental artefacts in CARS is enhanced by using interpolation techniques for wavelets.
  The method is validated using CARS spectra of adenosine mono-, di-, and triphosphate in water, as well as, equimolar aqueous solutions of D-fructose, D-glucose, and their disaccharide combination sucrose.
\end{abstract}
\section{Introduction}
Coherent anti-Stokes Raman scattering (CARS) spectroscopy offers a unique microscopic tool in biophysics, biology, and materials research.
\cite{Muller:2002, Evans:2008, Min:2011, Garbacik:2011, Fussel:2014, Cheng:2015, Cleff:2016, Osseiran:17, Geissler:2017, Hirose:2018, Karuna:2019, Levchenko:2019, Nuriya:2019, Nishiyama:2020}
In addition to being ideally suited for qualitatively label-free microscopy \cite{Evans:2008, Min:2011, Cheng:2015}, the multiplex approach of
CARS can also provide complete (position-dependent) vibrational spectra. In principle, this would allow a quantitative, local analysis of chemical
composition \cite{Muller:2002, Rinia:2006, Muller:2007, Rinia:2008, Day:2011, Karuna:2019, Nuriya:2019}.
However, unlike a spontaneous Raman scattering spectrum, a CARS measurement as such does not directly provide any quantitative information.

An observed CARS spectrum arises from a coherent addition of both resonant contributions from different vibrational modes and a constant, non-resonant (NR) background contribution.
This results in
a complex line shape, where the positions, amplitudes, and line widths of each vibrational mode are generally hidden.
This is particularly true for condensed-phase samples, where the vibrational spectra are highly congested with strongly-overlapping vibrational modes \cite{Vartiainen:06}.
At a minimum, quantitative analysis requires extracting the Raman line shapes from CARS spectra.
This can be done by using a suitable phase retrieval method \cite{Vartiainen:06, Liu:09} on the normalized CARS spectrum.
However, the technology is still limited in terms of comparable and quantitative analysis methods, which remain active and ongoing topics of research. \cite{Vartiainen:06, Day:2011, Camp:2016, Kan:16}
Moreover, the analysis is complicated by experimental errors encountered in obtaining a normalized CARS line-shape spectrum, which leads to an erroneous, non-additive, and non-constant background
component to the NR background \cite{Camp:2016, Kan:16}.
If it remains uncorrected, this artefact in the NR background can prevent any
quantitative information from being obtained from a CARS measurement. Recently, a procedure based on the wavelet prism decomposition algorithm was proposed to address this issue \cite{Kan:16}.

Sequential Monte Carlo (SMC) methods have been successfully applied in a wide variety of contexts, including motion tracking\cite{Ababsa:2011,Liu:2015}, satellite image analysis \cite{Moores:2015}, medical applications \cite{Lee:2017}, and geophysics \cite{Leeuwen:2009}.
In spectroscopy, Bayesian methods such as SMC have recently been gaining significant attention from the research community.
Bayesian statistical inference has been applied to electrochemical impedance \cite{Effat:2017}, double electron-electron resonance \cite{Edwards:2016}, time-resolved analysis of gamma-ray bursts \cite{Yu:2019}, and in estimation of elastic and crystallographic features by resonance ultrasound spectroscopy \cite{Bales:2018} to name a few.
In particular, a hierarchical Bayesian approach combining modelling of individual line shapes with a continuous background model, with estimation done via SMC methods, has been introduced for Raman spectroscopy\cite{Moores:16}.

The contributions of this study are three-fold.
We introduce a method for correcting experimental artefacts in raw CARS measurements, extending further the existing method based on wavelet prism decomposition\cite{Kan:16}.
Secondly, we propose a line-narrowing method with improved properties compared to the Line Shape Optimized Maximum Entropy Linear Prediction (LOMEP) method\cite{Kauppinen:81, Kauppinen:91}.
Our method utilizes linear prediction, as in LOMEP, but in contrast circumvents the need of assuming a single \textit{a priori} common line shape for all spectral lines.
This constitutes a major improvement over the LOMEP method.
Thirdly and foremost, Bayesian inference is introduced to CARS spectrum analysis, extending previously available analysis methods.
We formulate a Bayesian inference model that is capable of estimating predictive distributions of the underlying Raman signal, error-corrected CARS spectrum, and the measurement CARS spectrum.
This is enabled by parametric modelling of Voigt line shapes, along with a continuous, wavelet-based model for experimental artefacts.

In what follows, we introduce the Bayesian statistical model for CARS.
The Raman signal of the CARS spectrum is modelled using a linear combination of Voigt line shapes. Using the Hilbert transform, we construct the modulus of the resonant part of the CARS spectrum.
A non-resonant part, estimated from the data\cite{Kan:16}, is added to obtain an error-free CARS spectrum, which is finally modulated with a slowly-varying error function.
Next, we describe the numerical algorithms used for statistical inference and line narrowing, along with our Bayesian prior distributions.
We then present experimental details along with obtained results for the predictive intervals for the resonant Raman signal and the constituent line shapes, modulating error function, error-corrected CARS spectrum, and the measurement CARS spectrum.
Lastly, the key aspects of the study are briefly remarked upon, thereby concluding the paper.

\section{Methods}
\subsection{Statistical Model}

We model CARS spectral measurements with an additive error model given as
\begin{align}
    y_k:=y(\nu_k) = f(\nu_k; p,\bm\theta) + \epsilon(\nu_k),
    \label{eq:observation}
\end{align}
where $y_k$ denotes a measurement that has been discretized with spectral sampling resolution $h>0$ at a wavenumber location $\nu_k=kh$ with $k\in\mathbb{Z}_+$, $f(\nu_k; p,\bm\theta)$ is the CARS spectrum model with parameter $p$ controlling the baseline and parameters $\bm\theta$ for the Voigt line shape, and with measurement error $\epsilon(\nu_k) \sim \mathcal{N}(0, \sigma_\epsilon^2)$ with known variance.
For the spectrum, we use a parameter-wise separable model
\begin{align}
    f(\nu; p,\bm\theta) = \varepsilon_{\rm{m}}(\nu; p)S( \nu; \bm\theta),
    \label{eq:dataModel}
\end{align}
where $p$ is the interpolated discrete wavelet transform (DWT) detail level, $\varepsilon_{\rm{m}}(\nu; p)$ is the modulating error function,
and $S( \nu; \bm\theta)$ is the error-corrected CARS signal, similar to the representation used in Ref.~\citenum{Kan:16}.
The signal $S$ can further be represented as
\begin{equation}\label{eq:signal}
\begin{split}
    S( \nu; \bm\theta) &= \left\vert\chi^{(3)}_{\mathrm{NR}}(\nu) + \chi^{(3)}_{\text{R}}(\nu; \bm\theta)\right\vert^2 \\
    &=\left\vert\exp\left( \frac{A_J(\nu)}{2}  \right)+ \left( iV_N( \nu; \bm\theta)-\mathcal{H}\{V_N( \nu; \bm\theta)\}\right)\right\vert^2,
    \end{split}
\end{equation}
where the exponential part corresponds to the non-Raman part with $A_J$ practically constant (see Ref.~\citenum{Kan:16} for details), $\mathcal{H}$ is the Hilbert transform, and
\begin{equation}
\begin{split}
    V_N( \nu; \bm\theta) &= \sum_{n=1}^N a_n V(\nu - \nu_n; \sigma_n, \gamma_n) = \sum_{n=1}^N a_n L(\nu - \nu_n; \gamma_n) \ast G(\nu - \nu_n; \sigma_n)\label{eq:raman}\\
    &= \sum_{n=1}^N a_n \frac{1}{\sqrt{2\pi \sigma_n^2}} \exp\left( - \frac{( \nu - \nu_n )^2}{2\sigma_n^2} \right)
    *
     \frac{1}{\pi \gamma_n}\frac{\gamma_n^2}{ ( \nu - \nu_n )^2 + \gamma_n^2},
\end{split}
\end{equation}
where $\ast$ denotes convolution.
$N$ stands for the number of line shapes, with each line shape having $\bm\theta_n:=(a_n,\nu_n,\sigma_n,\gamma_n)^T$ parameters standing for the amplitude, location, scale of the Gaussian shape, and scale of the Lorentzian shape, respectively.
Thus, we have $4 N$ parameters in total for our model of $S(\nu;\bm\theta)$.

Instead of the wavelet prism method \cite{Kan:16}, we model the modulating error function as
\begin{align}
    \log\big(\varepsilon_{\rm{m}}(\nu; p) \big) = \sum\limits_{j = {\ceil{p+1}}}^J D_j(\nu) + (1 - \beta)D_{\ceil{p}}(\nu),
    \label{eq:errorFunction}
\end{align}
where $p \in [1,J]$, and $\beta = p - \floor{p}$, i.e., as an interpolation between the discrete wavelet reconstruction levels $D_j$ to have a continuous model for the background.
With the above, we can have an unnormalized posterior formulated as
\begin{align}\label{eq:posterior}
    \pi\left( p, \bm\theta \mid \mathbf{y} \right) \propto \mathcal{L}\left( \mathbf{y} \mid p, \bm\theta\right) \pi_0\left( p, \bm\theta \right),
\end{align}
where $\mathbf{y} := (y_1, \dots, y_K)^T \in \mathbb{R}^K$ is the vector of observations given via eq \eqref{eq:observation}, $\bm\theta \in \mathbb{R}^{4N}_{+}$ is the parameter vector $(\bm\theta_1, \dots, \bm\theta_N)^T$ for the $N$ Voigt peaks, $\mathcal{L}( \mathbf{y} \mid p, \bm\theta)$ represents the likelihood distribution of the forward model, and $\pi_0( p, \bm\theta )$ denotes prior distributions for some or all of the model parameters $p$ and $\bm\theta$.
As such, the total number of parameters in the model is $4N + 1$. The solution of \eqref{eq:posterior} is unavailable in closed form, but following Ref.~\citenum{Moores:16} we can use Monte Carlo methods to obtain samples from this distribution, as described in the following section.

\subsection{Sequential Monte Carlo}
Sequential Monte Carlo (SMC) methods, also known as particle filtering and smoothing, are widely used in statistical signal processing \cite{Sarkka:2013}.
These algorithms provide a general procedure for sampling from Bayesian posterior distributions \cite{Chopin:2002, Moral:2006}.
SMC methods utilize a collection of weighted particles, initialized from a prior distribution, which are ultimately transformed to represent a posterior distribution under investigation.
The methodology used in this study is similar to the one used in Ref.~\citenum{Kan:16} where they use sequential likelihood tempering \cite{Moral:2006} to fit a model of surface-enhanced Raman spectra to measurements.

Assuming additive Gaussian measurement errors $\epsilon(\nu_k)$, likelihood of the model $f(\nu_k; p,\bm\theta)$ fitting measurement data $y_k$ can be formulated as
\begin{align}
    \mathcal{L}( \mathbf{y} \, | \,\, p, \bm\theta) \sim \prod\limits_{k = 1}^K \mathcal{N}( y_k; f(\nu_k; p, \bm\theta), \sigma_\epsilon^2),
\label{eq:pf_likelihood}
\end{align}
and the posterior distribution for step $t$ of the sequential likelihood tempering is given by
\begin{align}
    \pi^{(t)}( p, \bm\theta \, | \,\, \mathbf{y} ) \propto \mathcal{L}( \mathbf{y} \, | \,\, p, \bm\theta)^{\kappa^{(t)}}\pi_0( p, \bm\theta ),
    \label{eq:tempered_posterior}
\end{align}
where the superscript $(t)$ denotes the iteration or ``time" step of the algorithm and $\kappa^{(t)}$, $\kappa^{(t - 1)} < \kappa^{(t)} < \kappa^{(t + 1)} < \dots \leq 1$ with $\kappa^{(0)} = 0$, being a parameter controlling the degree of tempering of the likelihood, with the initial state being equal to the prior distribution while increasingly tempering the total likelihood towards the complete Bayes' theorem. The tempering parameter $\kappa^{(t)}$ can be defined simply as an strictly increasing sequence so that $\kappa^{(t)} \in [0, 1]$ or as done in Ref.~\citenum{Moores:16}, the parameter can be determined adaptively according to a given learning rate $\eta$ such that the relative reduction in the ESS between iterations is approximately $\eta$.

Using $Q$ particles, individual weights of each particle at initial step $t = 0$ are set as equally important $w_q^{(0)} = \frac{1}{Q}$.
The weights are then updated, and normalized, at each step $t$ according to
\begin{align}
    w_q^{(t)} \propto \frac{\mathcal{L}( \mathbf{y} \, | \,\, p, \bm\theta)^{\kappa^{(t)}}}{\mathcal{L}( \mathbf{y} \, | \,\, p, \bm\theta)^{\kappa^{(t-1)}}} w_q^{(t - 1)}.
    \label{eq:weight_update}
\end{align}
However, updating the particle weights gradually impoverishes the sample distribution.
This degradation is measured by the effective sample size (ESS)
\begin{align}
    Q_{\text{ESS}}^{(t)} = \frac{1}{\sum\limits_{q = 1}^Q\left( w_q^{(t)} \right)^2}.
    \label{eq:ess}
\end{align}
To counteract this, the particles are resampled when the ESS has fell below a set threshold $Q_{\text{min}}$ according to a chosen resampling algorithm and the particle weights are reset as $w_q^{(t)} = \frac{1}{Q}$.
Due to resampling, duplicates of the particles are obviously generated.
This is undesirable and as such, each particle is additionally updated using Markov chain Monte Carlo (MCMC) with the target distribution given by the tempered posterior distribution defined in eq \eqref{eq:tempered_posterior} at the current iteration or ``time" step $t$.
A pseudo-code of the SMC algorithm used is presented in Algorithm \ref{alg:particle_filter}.
\begin{algorithm}[H]
\caption{A sequential Monte Carlo sampler.}
\label{alg:particle_filter}
\begin{algorithmic}
\State \textbf{Initialize:}
	\State \hspace{\algorithmicindent} Set $t = 0$ and $\kappa^{(t)} = 0$.
	\State \hspace{\algorithmicindent} Draw $Q$ particles from the prior distribution $\pi_0( p, \bm\theta )$.
    \State \hspace{\algorithmicindent} Set particle weights $w_q = \frac{1}{Q}$.
\While{$\kappa_t < 1$}
    \State $t = t + 1$.
    \State Determine $\kappa^{(t)}$.
    \State Update particle weights $w_q^{(t)}$ according to \eqref{eq:weight_update}.
    \State Compute the effective sample size $Q_{\text{ESS}}^{(t)}$ using \eqref{eq:ess}.
    \If{$Q_{\text{ESS}}^{(t)} < Q_{\text{min}}$}
    	\State Resample particles according to their weights.
    	\State Set particle weights $w_q = \frac{1}{Q}$.
    \EndIf
    \State Update the particles with MCMC using the tempered posterior given by \eqref{eq:tempered_posterior}.
    \State Update particle weights $w_q^{(t)}$ according to their likelihoods.
    \State Recompute the effective sample size $Q_{\text{ESS}}^{(t)}$ using \eqref{eq:ess}.
\EndWhile
\end{algorithmic}
\end{algorithm}
\subsection{Line Narrowing Method}
In line narrowing one aims at spectral line sharpening.
We shall use line narrowing for making rough initial estimation of peak locations $\nu_n$, amplitudes   $a_n$, and number of line shapes $N$.
This is a preprocessing step for the statistical estimation method described in the previous section.
Let us model and approximate a spectrum with Lorentzian line shapes as
\begin{equation} \label{eq:assumptionOfData}
	\widetilde{V}_N(\nu_k, \widetilde{\bm\theta}) := \sum\limits_{n=1}^N a_nL(\nu_k; \nu_n, \gamma_n) \approx \sum\limits_{n=1}^N a_nL(\nu_k; \nu_n, \gamma)
    = L(\nu_k; 0, \gamma) \ast \sum\limits_{n=1}^N a_n\delta(\nu_k - \nu_n),
\end{equation}
where $\widetilde{V}_N(\nu_k, \widetilde{\bm\theta})$ denotes a measured spectrum, $\nu_k \in \mathbb{R}^K$ measurement locations, $\widetilde{\bm\theta} := \linebreak (a_n, \nu_n, \gamma_n)^T$, $\gamma$ a single constant width parameter, and $\delta(\nu - \nu_n)$ is the Dirac delta function.

Our starting point is the LOMEP method \cite{Kauppinen:81, Kauppinen:91}, where the constant $\gamma$ approximation is used.
With suitably chosen $\gamma$, we have
\begin{align}
     \mathcal{F} \left\{\sum\limits_{n=1}^N a_n\delta(\nu - \nu_n)\right\} = \frac{\mathcal{F} \left\{\widetilde{V}_N(\nu_k, \widetilde{\bm\theta}) ) \right\}}{\mathcal{F}\left\{L(\nu_k; 0, \gamma)\right\}} =: x_\textrm{LP}(t_k;\gamma, N_\textrm{FIR})  ,
\end{align}
where $\mathcal{F}$ denotes the Fourier transform, $t_k$ the Fourier domain variable, and $x_\textrm{LP}(t_k;\gamma_m, N_\textrm{FIR})$ is the linearly predicted time signal.
In LOMEP, the linear prediction is done finite impulse response filtering with filter length $N_\textrm{FIR}-1$\cite{Kauppinen:81, Kauppinen:91}.
The major limitation of LOMEP is the heuristic choice of $\gamma$, and additionally, the $q$-curve optimization method fails when  number of line shapes $N$ increases.
Despite the drawbacks, the potential of the linear prediction scheme is nevertheless attractive for its ability to heavily narrow down line shapes when successful.

We propose an alternative approximative model in eq \eqref{eq:assumptionOfData} as a linear combination of $M$ similarly constructed convolutions
\begin{equation}
    \sum\limits_{n=1}^N a_nL(\nu; \nu_n, \gamma_n)  \approx \frac{1}{M}\sum\limits_{m=1}^M \sum\limits_{n=1}^N a_nL(\nu; \nu_n, \gamma_m)
    = \frac{1}{M}\sum\limits_{m=1}^M \left( L(\nu; 0, \gamma_m) \ast \sum\limits_{n=1}^N a_n\delta(\nu - \nu_n) \right),
    \label{eq:lineNarrowingModel}
\end{equation}
using a set of width parameters $\gamma_m$ in contrast to fixed $\gamma$.
Then, the approximation of the Dirac delta functions is
\begin{align}
    D_\textrm{A}(\nu_k, \gamma_m, N_\textrm{FIR}) = \mathcal{F}^{-1}\left\{ x_\textrm{LP}(t_k;\gamma_m, N_\textrm{FIR}) \right\}  \approx \sum\limits_{n=1}^N a_n\delta(\nu_k - \nu_n).
\end{align}
The squared sum of residuals for a single convolution, denoted here by $d(\gamma_m, N_\textrm{FIR})$, can be given as
\begin{align}
    d(\gamma_m, N_\textrm{FIR}) = \big\Vert \widetilde{V}_N(\nu_k, \widetilde{\bm\theta}) - L(\nu_k; 0, \gamma_m) \ast D_\textrm{A}(\nu_k, \gamma_m, N_\textrm{FIR}) \big\Vert_2^2.
    \label{eq:sumOfResiduals}
\end{align}
We additionally define a constrained squared sum of residuals as
\begin{equation}
\begin{split}
    d_{\textrm{C}}(\gamma_m, N_\textrm{FIR}) = \left\Vert \widetilde{V}_N(\nu_k, \widetilde{\bm\theta}) - c_\textrm{n} L(\nu_k; 0, \gamma_m) \ast \mathbf{1}_{D_\textrm{A} >0} D_\textrm{A}(\nu_k, \gamma_m, N_\textrm{FIR})  \right\Vert_2^2,
    \label{eq:regularizedSumOfResiduals}
\end{split}
\end{equation}
where $\mathbf{1}_{D_\textrm{A} >0}=1$, if $D_\textrm{A} >0$ and $0$ otherwise, and $c_\textrm{n}$ is a normalization constant so that the area under the spectrum is conserved:
\begin{align}
	c_\textrm{n}  &=  \ddfrac{ \sum\limits_{k=1}^K D_\textrm{A}(\nu_k, \gamma_m, N_\textrm{FIR}) }{ \sum\limits_{k=1}^K \mathbf{1}_{D_\textrm{A} >0} D_\textrm{A}(\nu_k, \gamma_m, N_\textrm{FIR}) }.
\end{align}
With $d_{\textrm{C}}(\gamma_m, N_\textrm{FIR})$ we truncate any negative parts of $D_\textrm{A}(\nu_k, \gamma_m, N_\textrm{FIR})$ and distort the truncated spectrum according to the normalization constant $c_\textrm{n}$ the more signal energy is present on the negative parts.
By Parseval's theorem, and by using an orthonormal wavelet basis, the energy of a signal $g(\nu)$ can be represented as
\begin{align}
\int\limits_{-\infty}^\infty | g(\nu) |^2 \, \text{d}t &= \sum_{l=-\infty}^{\infty} | a(l) |^2 + \sum_{j=0}^{\infty} \sum_{\kappa=-\infty}^{\infty} | b_j(\kappa) |^2
\end{align}
where $a$ and $b$ are the scaling function and wavelet coefficients obtained using DWT.
Given a signal with sharp features, the energy of the signal should be concentrated on the wavelet coefficients $b_j$ and, a measure of this concentration of wavelet coefficient energy (we) can be defined as
\begin{align}
C_{\textrm{we}} = \ddfrac{\sum_{j=0}^{\infty} \sum_{\kappa=-\infty}^{\infty} | b_j(\kappa) |^2}{\sum_{l=-\infty}^{\infty} | a(l) |^2 + \sum_{j=0}^{\infty} \sum_{\kappa=-\infty}^{\infty} | b_j(\kappa) |^2}.
\end{align}
With the above formulations, we propose Algorithm \ref{alg:line_narrowing}:
Define a set of width parameters $\gamma_m$, for example, inferred from computational chemistry.
Similarly, define an upper bound for the impulse response parameter $N_\textrm{FIR}$.
Then, compute $D_\textrm{A}(\nu_k, \gamma_m, N_\textrm{FIR})$ using linear prediction for all parameter combinations of $\gamma_m$ and $N_\textrm{FIR}$ and residuals $d$ and $d_{\textrm{C}}$ along with the wavelet energy concentrations $C_{\textrm{we}}$.

Then using the filtering criterion $f_\textrm{c} = d + d_{\textrm{C}}$, narrow down the set of possible solutions by sorting them according to $f_\textrm{c}$ and $C_{\textrm{we}}$.
Take a percentage $p_\textrm{we}$ of the wavelet energy sorted solutions, including the largest energy concentrations.
Similarly, take a percentage $p_{f_\textrm{c}}$ of the filtering criterion sorted solutions, including the smallest filtering criteria.
Thus, an intersection of these sets should include solutions with mostly positive and sharp line shapes.
Sort this intersection set of size $\widetilde{M}$ according to $d$.
Finally, estimate eq \eqref{eq:lineNarrowingModel} by choosing $M$ so that the sum of residuals $d_M$ is minimized:
\begin{align}
    \argmin_{M \leq \widetilde{M}}  d_M = \argmin_{M \leq \widetilde{M}} \left\Vert \widetilde{V}_N(\nu_k, \widetilde{\bm\theta}) - \frac{1}{M}\sum\limits_{m=1}^M L(\nu; 0, \gamma_m) \ast D_\textrm{A}(\nu_k, \gamma_m, N_\textrm{FIR}) \right\Vert_2^2.
    \label{eq:residuals_dm}
\end{align}
As needed, smooth the obtained line narrowed spectrum with a smoothing function.
\begin{algorithm}[H]
\caption{Line narrowing algorithm.}
\label{alg:line_narrowing}
\begin{algorithmic}
\State \textbf{Initialize:}
	\State \hspace{\algorithmicindent} Set $\gamma_m$.
	\State \hspace{\algorithmicindent} Set $N_\textrm{FIR}$.
\For{$\gamma_m$}
    \For{$N_\textrm{FIR}$}
        \State Apply linear prediction using $\gamma_m$ and $N_\textrm{FIR}$.
        \State Compute $d$, $d_{\textrm{C}}$, and $C_{\textrm{we}}$.
    \EndFor
\EndFor
\State \textbf{Construct the solution:}
\State \hspace{\algorithmicindent} Filter out a set of possible solutions according to $f_\textrm{c}$ and $C_{\textrm{we}}$.
\State \hspace{\algorithmicindent} Sort the possible solutions according to $d$.
\State \hspace{\algorithmicindent} Compute $d_M$ and choose the $M$ solutions which minimize $d_M$.
\State \textbf{Smoothing:}
\State \hspace{\algorithmicindent} Convolute the result using an appropriate smoothing kernel.
\end{algorithmic}
\end{algorithm}
\subsection{Priors}
We obtained priors by manually correcting for the experimental artefacts modelled by eq \eqref{eq:errorFunction} and simultaneously applying phase retrieval \cite{Vartiainen:92, Vartiainen:06, Liu:09, Cicerone:2012} and computation of the resonant imaginary component of the CARS spectrum until a reasonable Raman signal was observed.
The line narrowing algorithm was applied on the manually estimated Raman signal, producing a line narrowed spectrum from which individual line shapes could be identified.
We follow Ref.~\citenum{Moores:16} in setting informative priors for the line shape locations $\nu_k$ as normal distributions
\begin{align}
    \pi_0(\nu_n) = \mathcal{N}\left( \mu_{\nu_n}, \sigma_{\nu_n}^2\right),
    \label{eq:location_prior}
\end{align}
where $\mu_{\nu_n}$ and $\sigma_{\nu_n}^2$ are estimated for each line shape $V(\nu,\bm\theta_n)$ by numerically integrating perceived individual line shapes in the line narrowed spectrum to estimate the means $\mu_{\nu_n}$ and variances $\sigma_{\nu_n}^2$.
The line narrowing algorithm utilizes multiple Lorentzian line shapes with differing scale parameters $\gamma_m$, thereby giving access to an informative prior for $\gamma_n$.
As in Ref.~\citenum{Moores:16}, we set a prior common for each $\gamma_n$ as a log-normal distribution:
\begin{align}
    \pi_0\left(\log(\gamma_n)\right) = \mathcal{N}\left( \mu_{\log(\gamma)}, \sigma_{\log(\gamma)}^2\right),
\end{align}
where the estimates for the mean and variance, $\mu_{\log(\gamma)}$ and $\sigma_{\log(\gamma)}^2$, are obtained from the parameters contained in the intersection set of size $\widetilde{M}$.
Priors for the Gaussian shape parameters $\sigma_n$ are obtained by scaling $\pi_0\big(\log(\gamma_n)\big)$ by $\sqrt{2\log(2)}$.
This would correspond to using identical priors for the full-width at half maximums for both the Gaussian and Lorentzian line shapes.
For the amplitudes, we can obtain an estimate for the areas straight-forwardly by the same numerical integration used to estimate the priors for the locations, as described above.
We set a fairly wide prior for the amplitude by setting them as
\begin{align}
       \pi_0(a_n) = \mathcal{N}\left( \mu_{a_n}, \left(\frac{\mu_{a_n}}{4}\right)^2\right),
\end{align}
where the mean $\mu_{a_n}$ is the numerically integrated area of each line shape.
A prior for the background parameter $p$ is set as a uniform prior:
\begin{align}
    \pi_0(p) = \mathcal{U}\left(p_\text{min}, p_\text{max}\right).
\end{align}
An estimate for the noise level $\sigma_\epsilon^2$ was also obtained using the line narrowing algorithm.
The algorithm fits a smooth representation of the Raman spectrum to the manually corrected data according to eq \eqref{eq:residuals_dm}.
This smooth representation of the Raman signal is then transformed to the measurement space by eq \eqref{eq:signal} and then by eq \eqref{eq:dataModel}.
The resulting residuals between the transformed smooth Raman signal and the measured CARS spectrum were used as an estimate for the noise variance $\sigma_\epsilon^2$.
Detailed descriptions of priors specific for each experimental data set of fructose, glucose, sucrose, and adenosine phosphate can be found in the supplementary material.
\section{Experimental details}
\subsection{Samples}
The sugar samples used in the multiplex CARS spectroscopy were equimolar aqueous solutions of D-fructose, D-glucose, and their disaccharide combination sucrose.
For sample preparation D-fructose, D-glucose, and their disaccharide combination, sucrose ({$\alpha$}-D-glucopyranosyl-(1{$\to$}2)-{$\beta$}-D-fructofuranoside) were dissolved in buffer solutions (50 mM \linebreak HEPES, pH=7) at equal molar concentrations of 500 mM \cite{Muller:2007}.
The adenosine phosphate sample was an equimolar mixture of AMP, ADP and ATP in water for a total concentration of 500 mM \cite{Vartiainen:06}.
The adenine ring vibrations \cite{Mathlouthi:1980} are found at identical frequencies for either for AMP, ADP or ATP around 1350 cm$^{-1}$.
The phosphate vibrations between 900 and 1100 cm$^{-1}$ can be used to discriminate between the different nucleotides \cite{Rinia:2006}.
The tri-phosphate group of ATP shows a strong resonance at 1123 cm$^{-1}$, whereas the monophosphate resonance of AMP is found at 979 cm$^{-1}$.
For ADP a broadened resonance is found in between at 1100 cm$^{-1}$.

\subsection{Multiplex CARS Spectroscopy}

All CARS spectra were recorded using a multiplex CARS spectrometer, the detailed description of which can be found elsewhere \cite{Muller:2002, Rinia:2006}.
In brief, a 10-ps and an 80-fs mode-locked Ti:sapphire lasers were electronically synchronized and used to provide the narrowband pump/probe and broadband Stokes laser pulses in the multiplex CARS process.
The center wavelengths of the pump/probe and Stokes pulses were 710 nm.
The Stokes laser was tunable between 750 and 950 nm.
The sugar spectra were probed within a wavenumber range from 700 to 1250 cm$^{-1}$, and the AMP/ADP/ATP spectrum within a range from 900 to 1700 cm$^{-1}$.
The linear and parallel polarized pump/probe and Stokes beams were made collinear and focused with an achromatic lens into a tandem cuvette.
The latter could be translated perpendicular to the optical axis to perform measurements in either of its two compartments, providing a multiplex CARS spectrum of the sample and of a non-resonant reference under near identical experimental conditions.
Typical average powers used at the sample were 95 mW (75 mW, in case of AMT/ADP/ATP) and 25 mW (105 mW) for the pump/probe and Stokes laser, respectively.
The anti-Stokes signal was collected and collimated by a second achromatic lens in the forward-scattering geometry, spectrally filtered by short-pass and notch filters, and focused into a spectrometer equipped with a CCD camera.
The acquisition time per CARS spectrum was 200 ms for sugar spectra and 800 ms for the AMP/ADP/ATP spectrum.
\subsection{Computational Details}
The SMC algorithm was computed using $Q = 2000$ particles with the resampling threshold set to $Q_{\text{min}} = 1000$ and the learning parameter set as $\eta = 0.9$.
Resampling was done, as in Ref.~\citenum{Moores:16}, via residual resampling \cite{Douc:2005}.
Target MCMC acceptance rate was set to 0.23 and the number of MCMC updates at each iteration was 200.
An AMD Ryzen 3950X processor was used with 27 CPU threads utilized, with the SMC estimation taking 580, 522, 413, and 688 seconds to produce the final posterior estimate of the parameters for the fructose, glucose, sucrose, and phosphate samples respectively.
For modelling the modulating error function $\varepsilon_{\rm{m}}(\nu; p)$ symlet 34 basis functions were used.

The line narrowing algorithm was run with $\gamma_m \in [1, 35]$ linearly spaced using 33 points.
The maximum number of measurement points $N_\textrm{FIR}$ used was 150, meaning that $N_\textrm{FIR} = \{1,\dots, 150\}$.
The length of the extrapolated signal \cite{Kauppinen:81, Kauppinen:91} was set to equal the number of measurement points in each spectrum.
The percentages $p_\textrm{we}$ and $p_{f_\textrm{c}}$ were set as 50\% and 2.5\% respectively.
To ensure that $\widetilde{M} > 0$, $p_{f_\textrm{c}}$ was incrementally increased by 2.5\% until a minimum intersection set size $\widetilde{M} \geq 50$ was achieved.
For computation of the wavelet energy concentration $C_{\textrm{we}}$ symlet 8 basis functions were used.
\section{Results and discussion}
\begin{figure}[H]
    \centering
    \includegraphics[width=\textwidth]{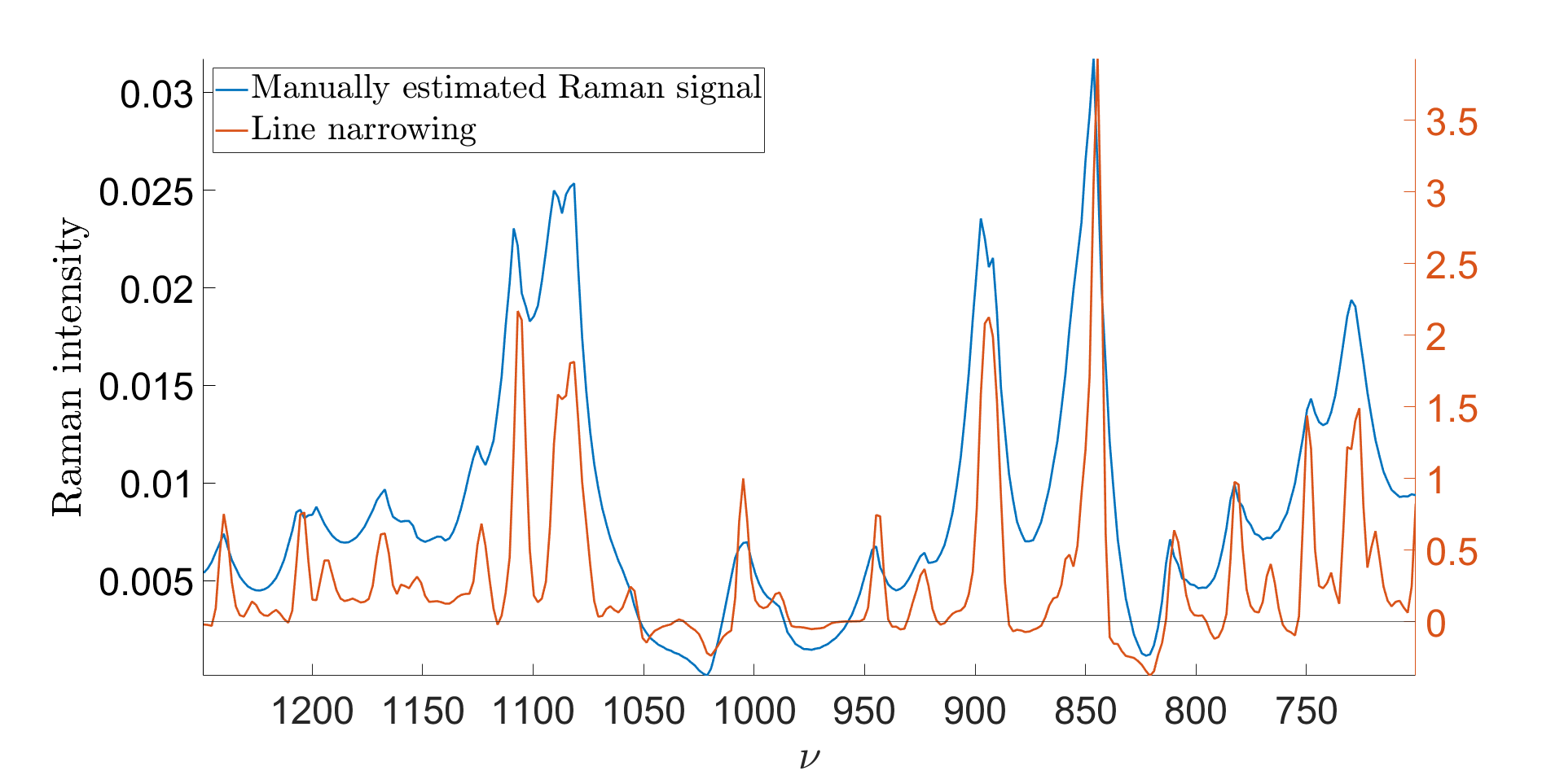}
    \caption{The manually estimated Raman signal of the fructose sample and the line narrowed Raman signal are shown in blue and red respectively. The perceivable individual line shapes were numerically integrated to yield informative prior estimates for Voigt line shape parameters.}
    \label{im:manualRamanFructose}
\end{figure}

\begin{figure}[H]
    \centering
    \subfloat[][]{\includegraphics[width=\textwidth]{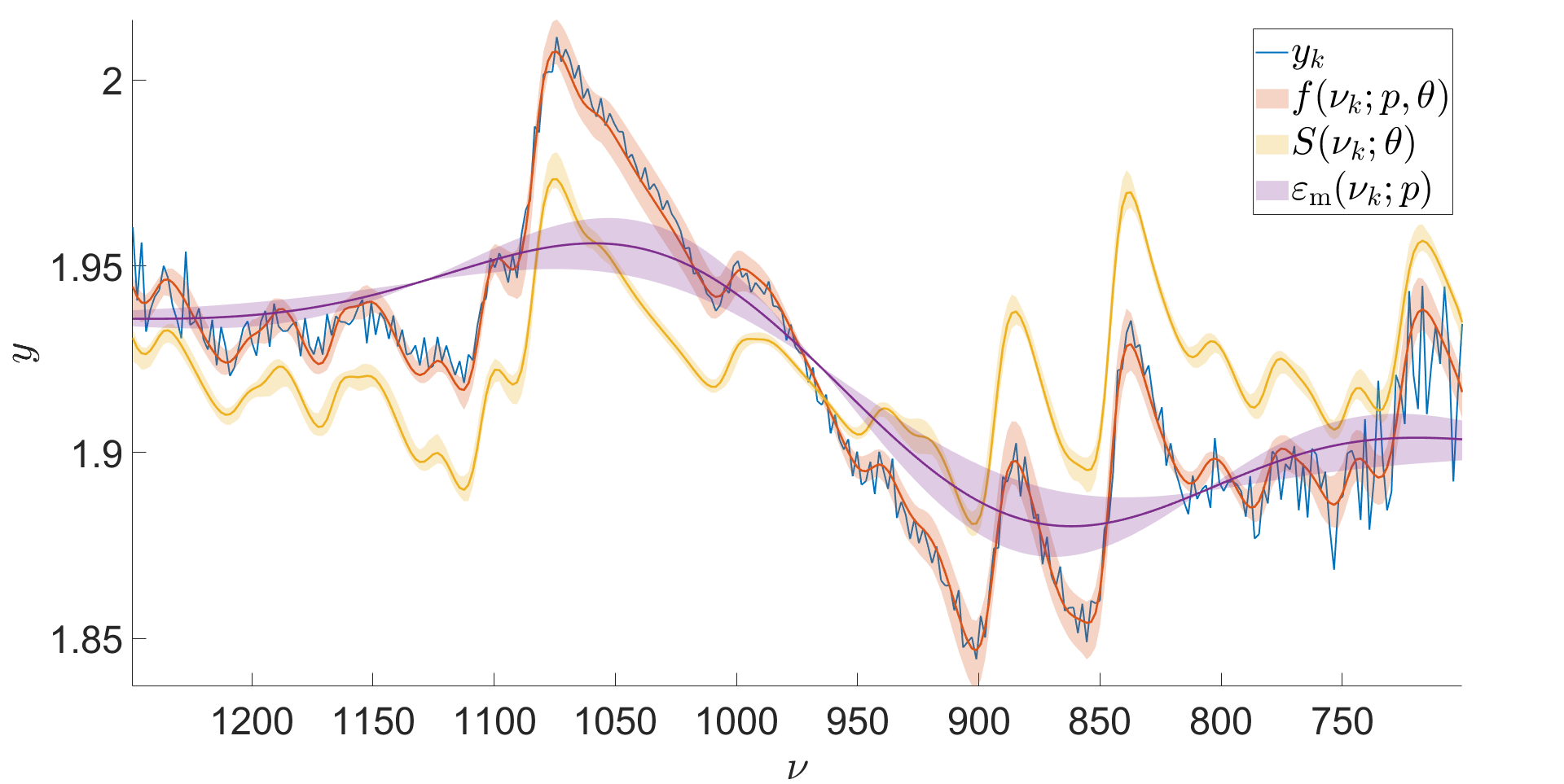}\label{im:fructosePredData}}\\
    \subfloat[][]{\includegraphics[width=\textwidth]{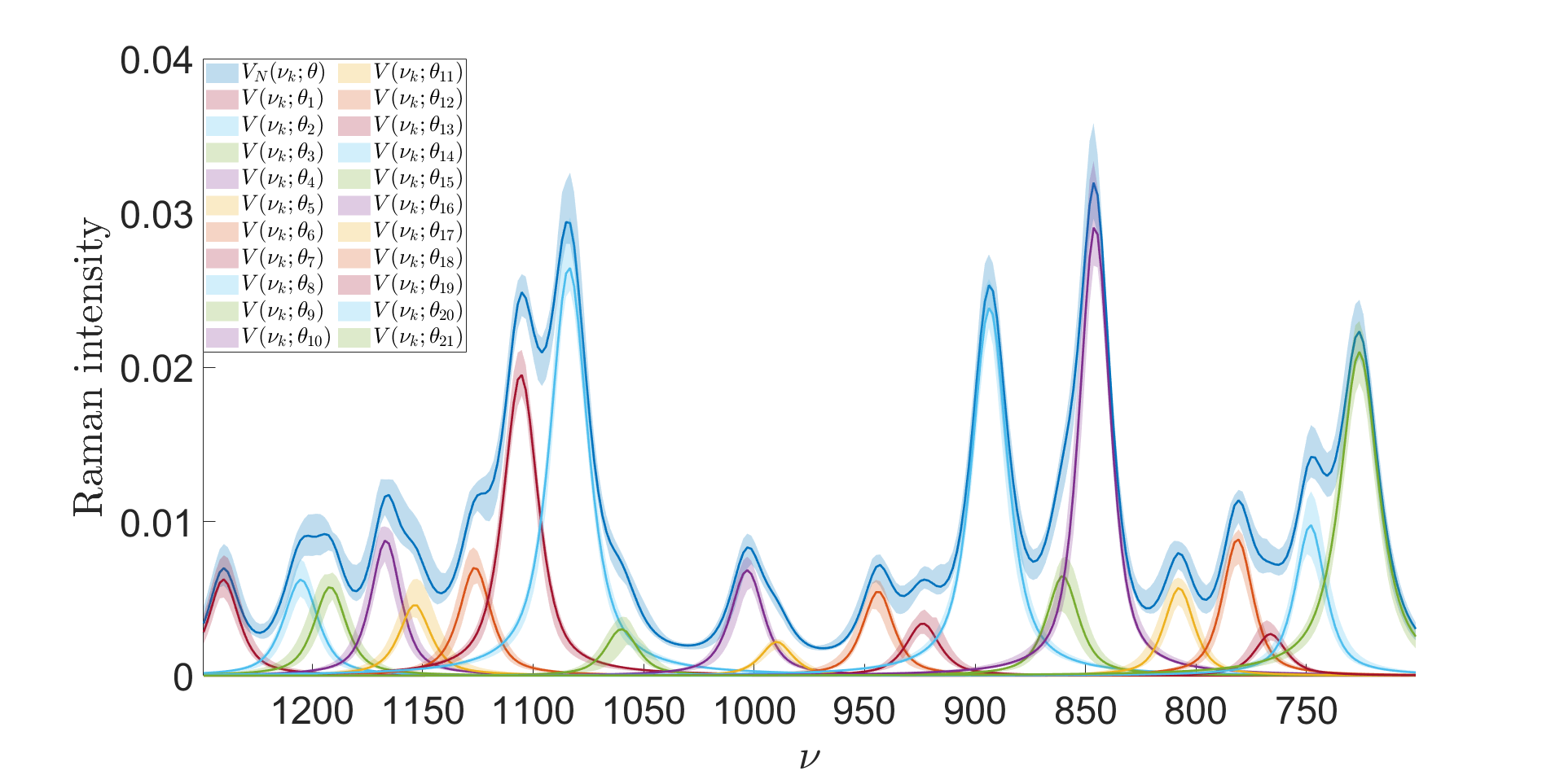}\label{im:fructosePredRaman}}
    \caption{In (a), the obtained 95\% predictive intervals for $y_k$, $f$, $S$, and $\varepsilon_{\rm{m}}$ shown in blue, red, yellow, and purple respectively for a CARS measurement of a fructose sample. In (b), the obtained 95\% predictive intervals for $V_N(\nu_k; \bm\theta)$ and each individual line shape $V(\nu_k; \bm\theta_n)$ for the fructose sample.}
\end{figure}

\begin{figure}[H]
    \centering
    \subfloat[][]{\includegraphics[width=\textwidth]{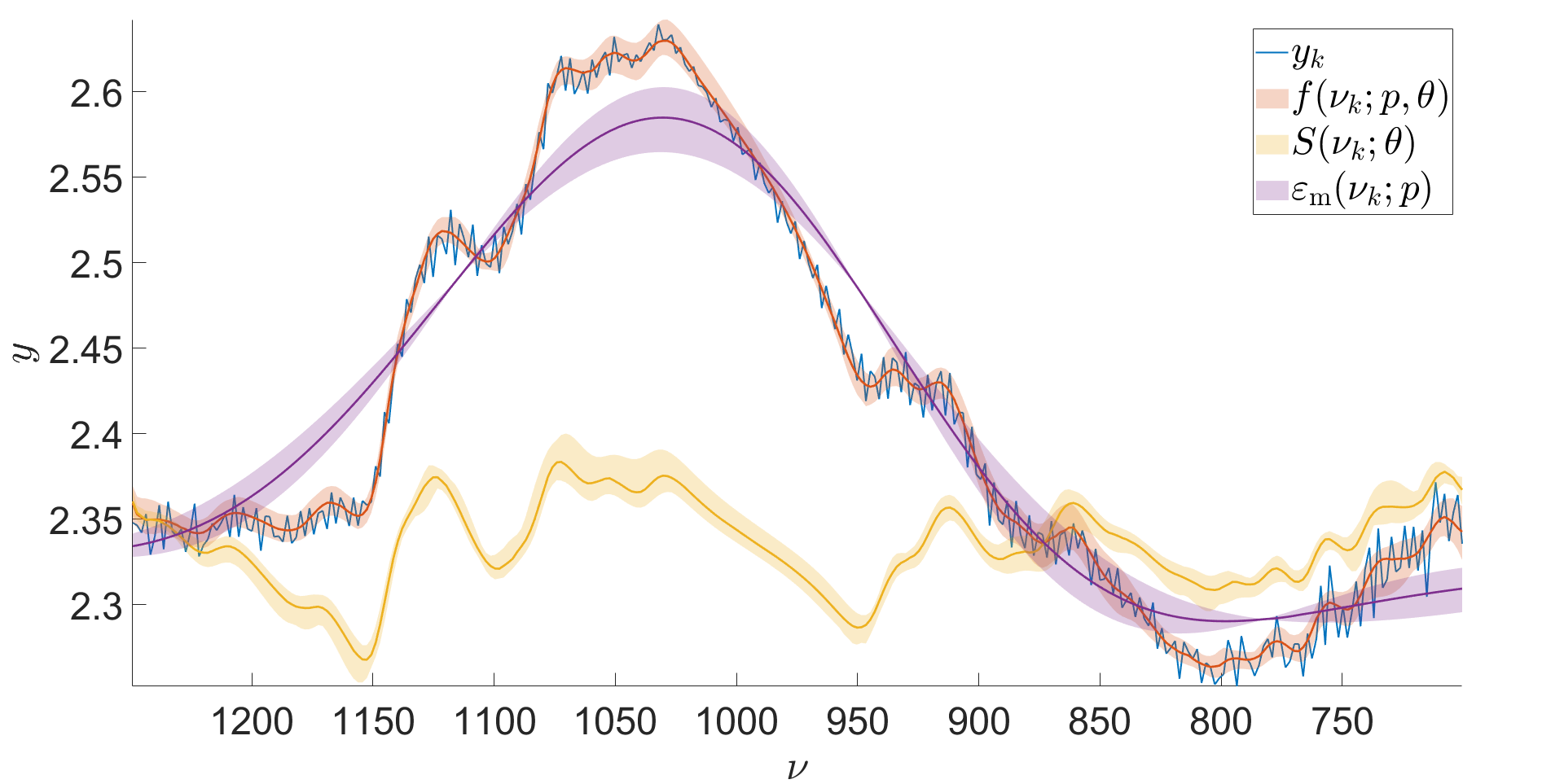}\label{im:glucosePredData}}\\
    \subfloat[][]{\includegraphics[width=\textwidth]{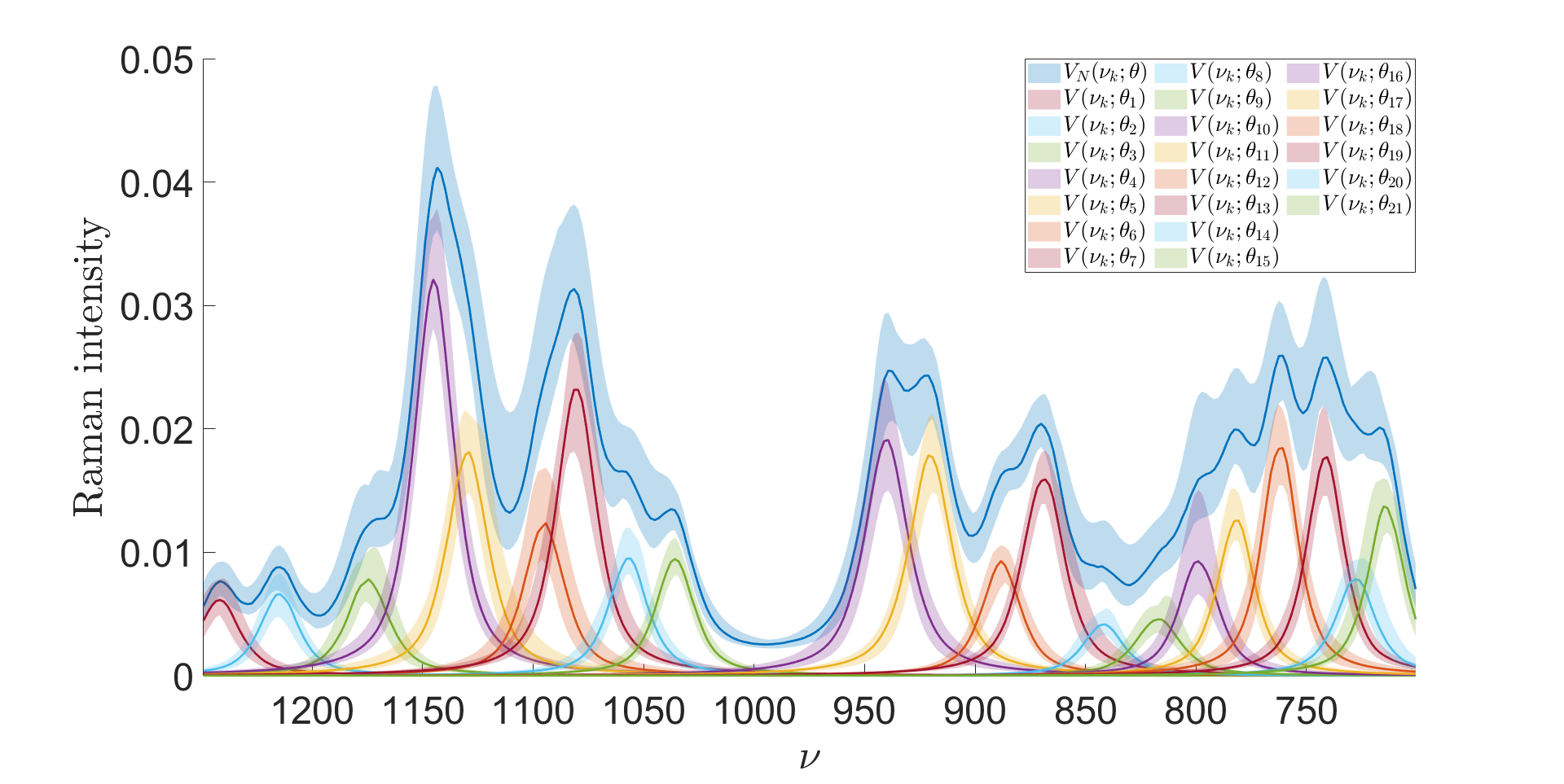}\label{im:glucosePredRaman}}
    \caption{In (a), the obtained 95\% predictive intervals for $y_k$, $f$, $S$, and $\varepsilon_{\rm{m}}$ shown in blue, red, yellow, and purple respectively for a CARS measurement of a glucose sample. In (b), the obtained 95\% predictive intervals for $V_N(\nu_k; \bm\theta)$ and each individual line shape $V(\nu_k; \bm\theta_n)$ for the glucose sample.}
\end{figure}

\begin{figure}[H]
    \centering
    \subfloat[][]{\includegraphics[width=\textwidth]{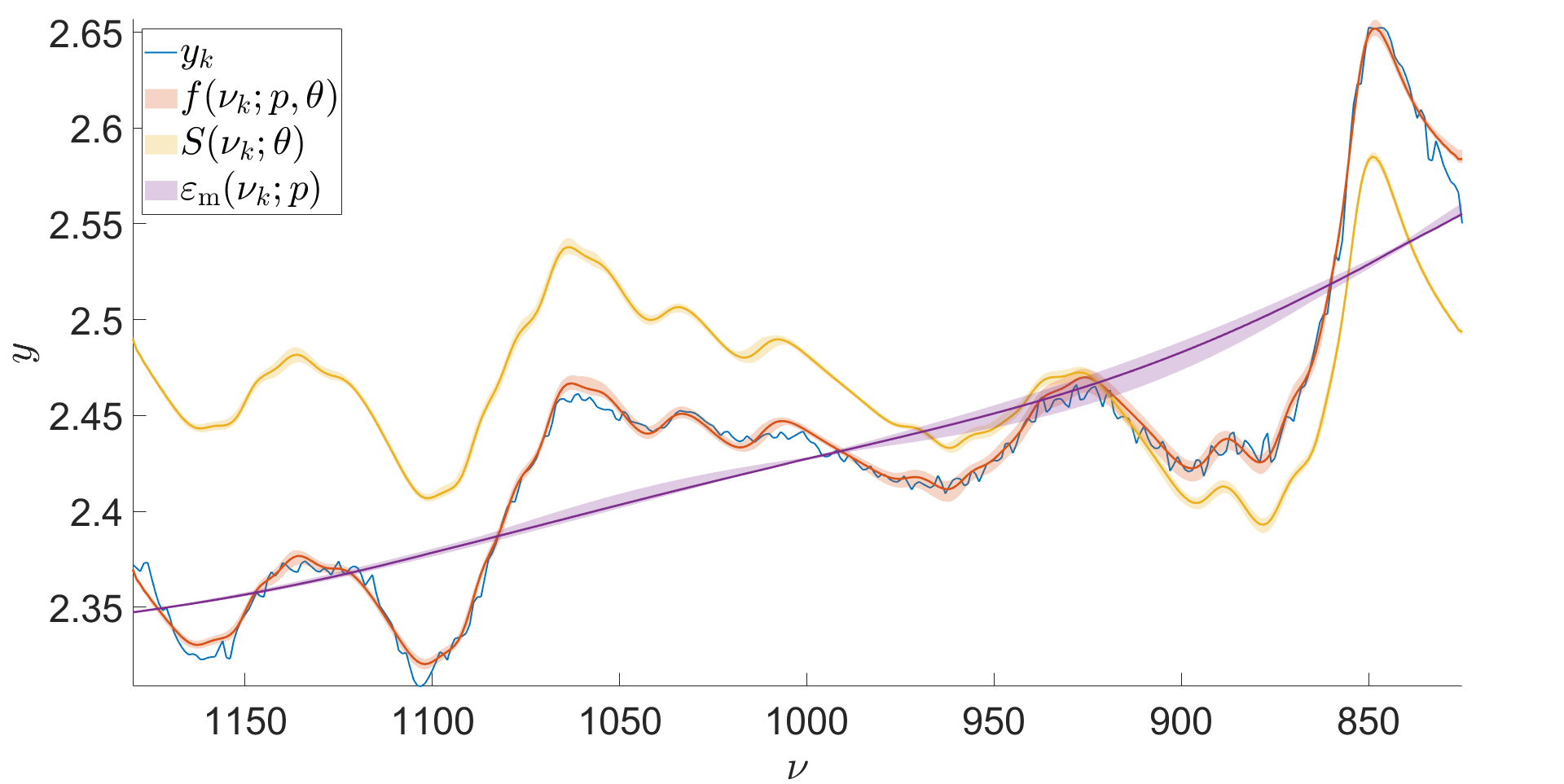}\label{im:predData}}\\
    \subfloat[][]{\includegraphics[width=\textwidth]{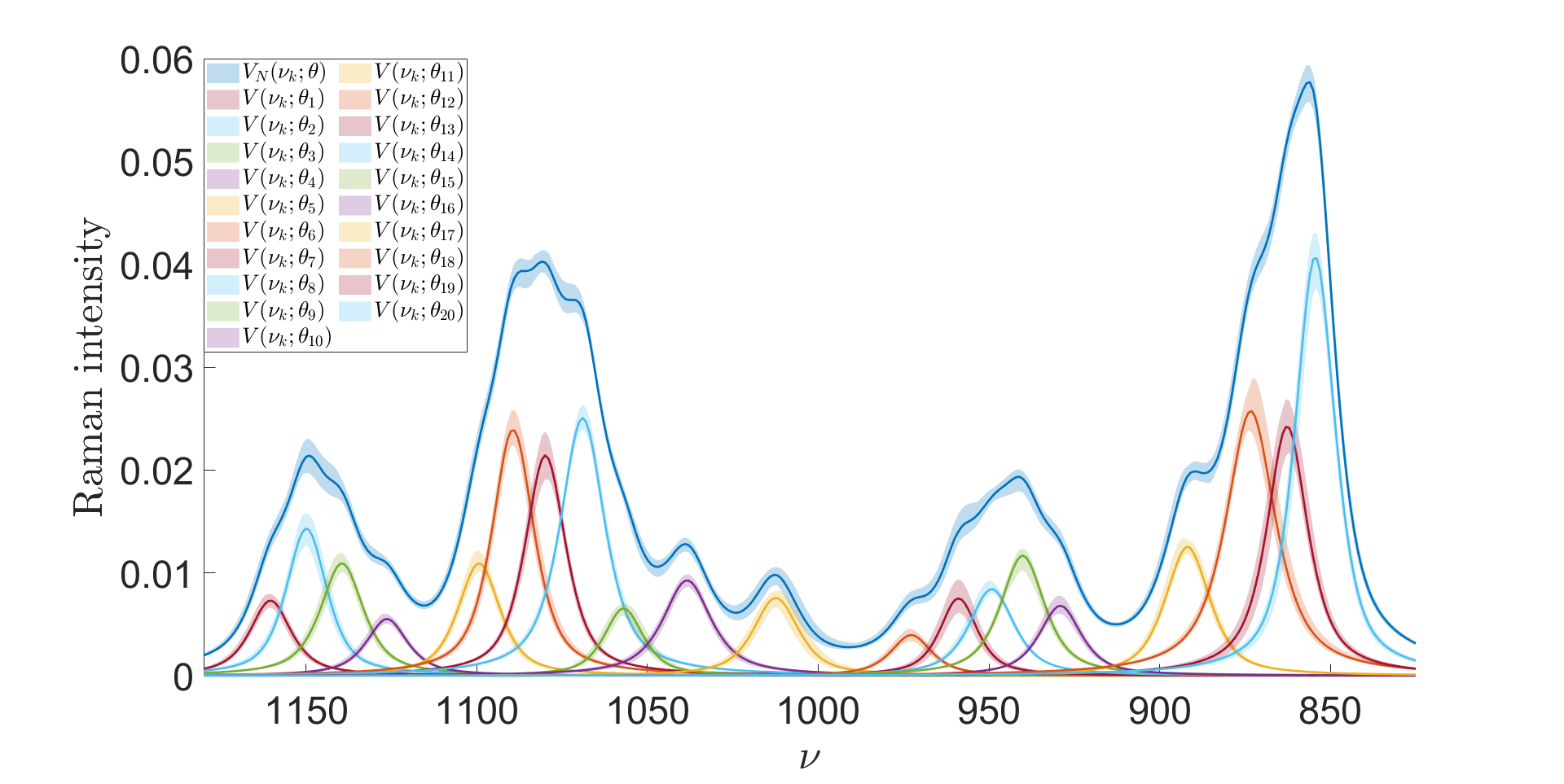}\label{im:sucrosePredRaman}}
    \caption{In (a), the obtained 95\% predictive intervals for $y_k$, $f$, $S$, and $\varepsilon_{\rm{m}}$ shown in blue, red, yellow, and purple respectively for a CARS measurement of a sucrose sample. Some discrepancies between $y_k$ and $f$ can be seen around the boundaries. These areas of the data should be ignored in the optimization. In (b), the obtained 95\% predictive intervals for $V_N(\nu_k; \bm\theta)$ and each individual line shape $V(\nu_k; \bm\theta_n)$ for the sucrose sample.}
\end{figure}

\begin{figure}[H]
    \centering
    \subfloat[][]{\includegraphics[width=\textwidth]{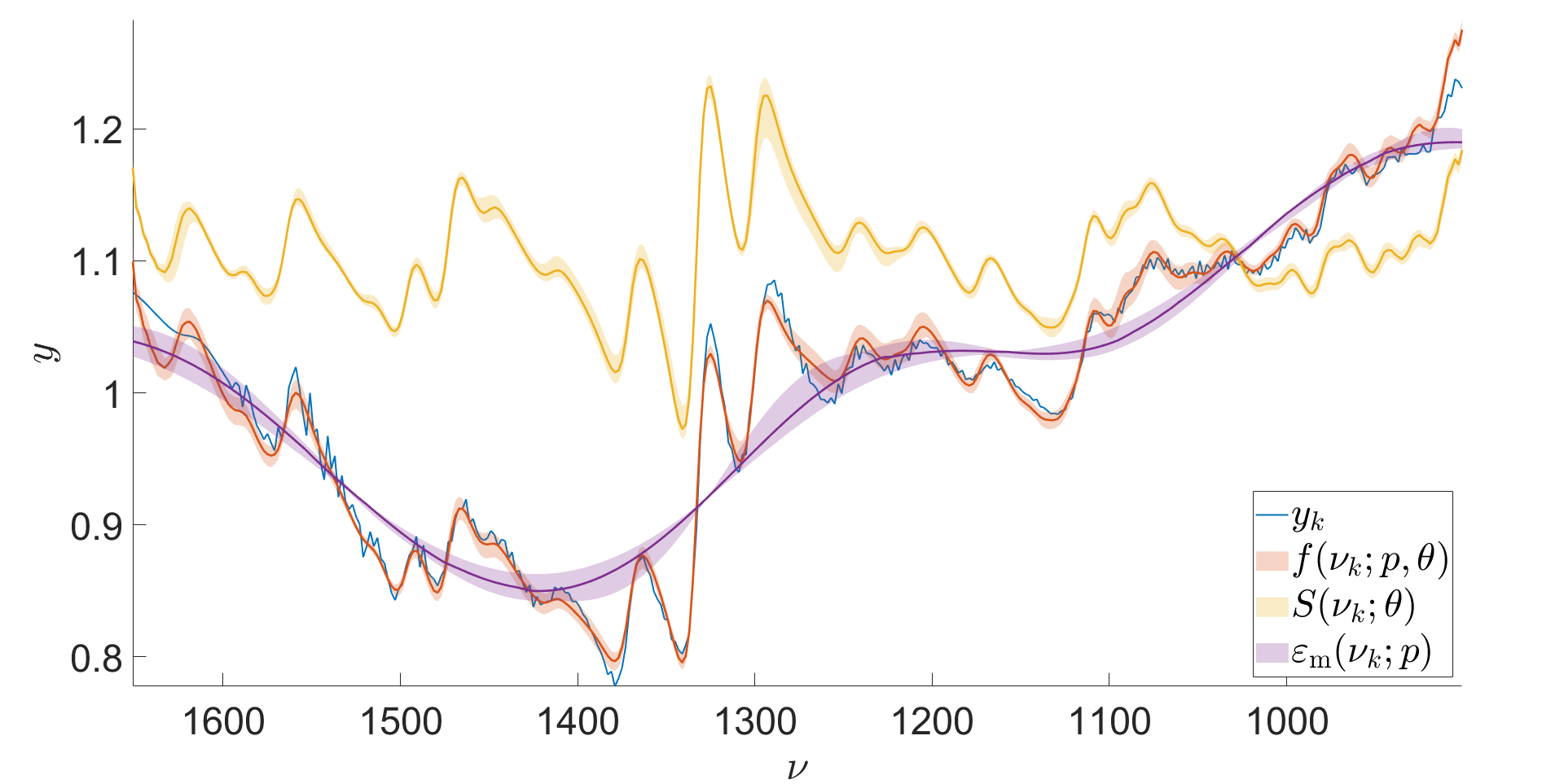}\label{im:phosphatePredData}}\\
    \subfloat[][]{\includegraphics[width=\textwidth]{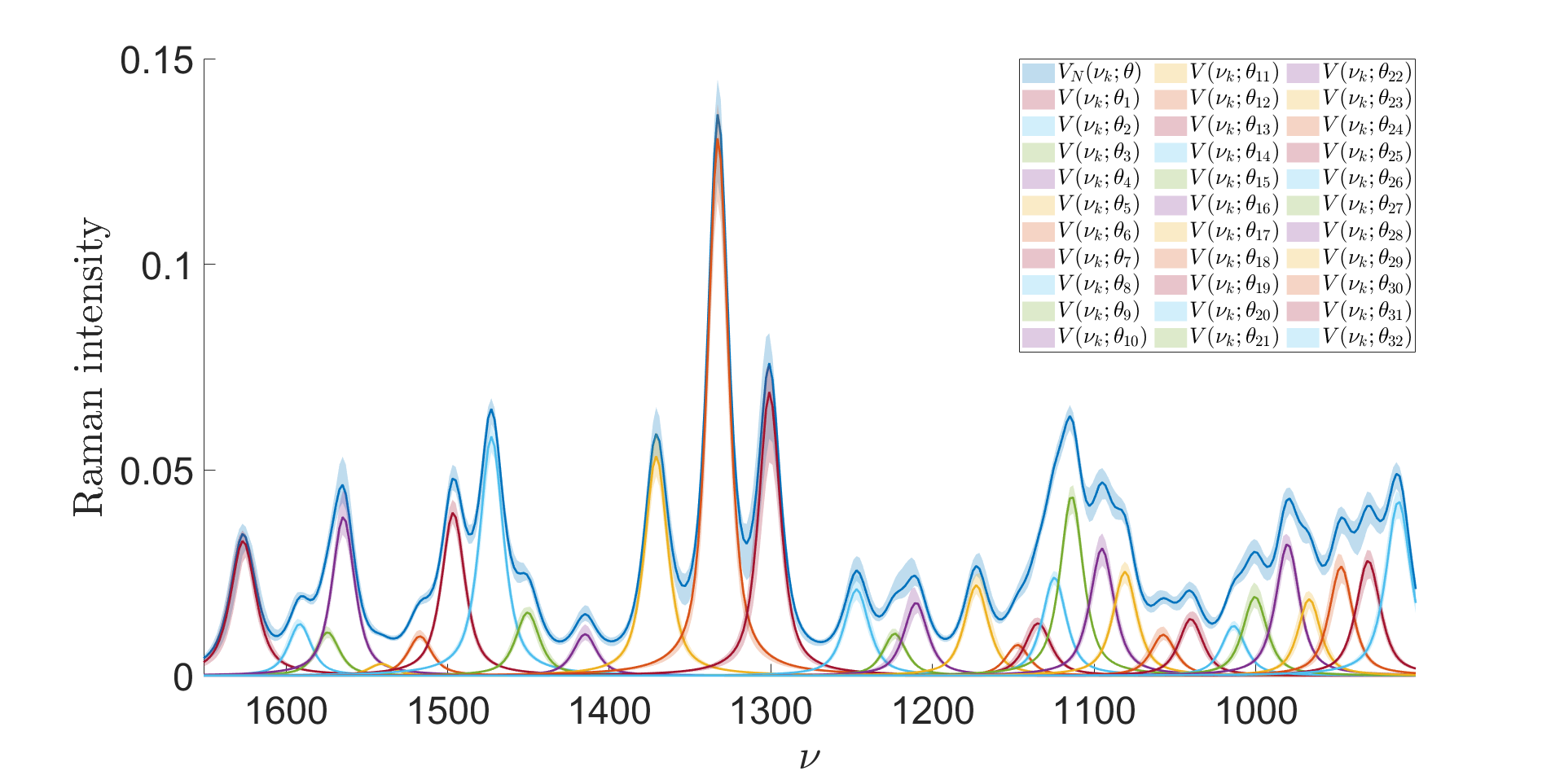}\label{im:phosphatePredRaman}}
    \caption{In (a), the obtained 95\% predictive intervals for $y_k$, $f$, $S$, and $\varepsilon_{\rm{m}}$ shown in blue, red, yellow, and purple respectively for a CARS measurement of a adenosine phosphate sample. In (b), the obtained 95\% predictive intervals for $V_N(\nu_k; \bm\theta)$ and each individual line shape $V(\nu_k; \bm\theta_n)$ for the adenosine phosphate sample.}
\end{figure}

In what follows, obtained 95\% predictive intervals for the forward model $f(\nu_k; p,\bm\theta)$, the modulating error function $\varepsilon_{\rm{m}}(\nu; p)$, and the error corrected spectra $S( \nu; \bm\theta)$ are presented in Figures \ref{im:fructosePredData}, \ref{im:glucosePredData}, \ref{im:predData}, and \ref{im:phosphatePredData} for the experimental spectra of fructose, glucose, sucrose, and phosphate respectively.
Similarly, in Figures \ref{im:fructosePredRaman}, \ref{im:glucosePredRaman}, \ref{im:sucrosePredRaman}, and \ref{im:phosphatePredRaman} the 95\% predictive interval for the Raman signal represented by $V_N(\nu, \bm\theta)$ is presented along with the predictive intervals for each constituent line shape $V(\nu,\bm\theta_n)$.
To illustrate how the priors were estimated, the manually corrected Raman signal and the result obtained via the proposed line narrowing method are shown in Figure \ref{im:manualRamanFructose}.
Additionally, the obtained posterior distributions for $\bm\theta$, alongside their respective prior distributions, are presented in the supplementary material.

The inference model proposed here was found to adequately model the CARS measurements along with perceived noise levels in the spectra.
For future work, it would be interesting to include heteroscedasticity in the model instead of assuming a constant measurement error variance.
Comparing the estimated predictive intervals of the obtained Raman signal showed clear correspondence to measured Raman intensities of aqueous solutions for fructose and glucose\cite{Soderholm:1999}.
For aqueous solution of sucrose, number of estimated line shapes was considered to resemble the 18 line shapes reported for solid sucrose \cite{Brizuela:2012}.
The estimated priors were considered not to restrict the parameter posterior which can be observed in the posterior distributions when seen alongside the respective priors, especially so for the cases of fructose, sucrose, and adenosine phosphate.
The obtained Raman signals for fructose, glucose, and sucrose are similar to results obtained Ref.~\citenum{Kan:16} which further supports the applicability of the methodology presented in this study.

Obtaining informative priors can be approached in different ways for chemically known samples as was done in Ref.~\citenum{Moores:16} where the authors use results obtained by density functional theory (DFT) software to derive estimates for the location priors and existing studies on structural properties of a known sample such as \cite{Soderholm:1999, Brizuela:2012}.
Naturally, any other forms of information on the underlying line shapes could just as well be used for the prior distributions.
Here we have considered estimating the priors purely from the data using a line narrowing algorithm requiring little to no \textit{a priori} information on the sample under study, ignoring the fact that this information would clearly be available \cite{Soderholm:1999}.
Additionally, it is known that the use of maximum entropy methods in improving spectral resolution can cause individual line shapes to split \cite{Kauppinen:92, Kauppinen:93}.
In the proposed line narrowing method, the averaging together multiple resolution enhanced spectra is postulated to possibly lessen the effect of this undesired spectral line splitting.
\newpage
\subsection{Fructose}
\subsection{Glucose}
\subsection{Sucrose}
\subsection{Adenosine phosphate}

\section{Conclusion}
A Bayesian inference model applicable to coherent anti-Stokes Raman spectroscopy is proposed and numerically implemented.
This work extends the current methodology of analyzing CARS spectra by introducing Bayesian inference in the field, enabling uncertainty quantification of spectral features.
The statistical inference model is able to produce posterior distributions for physically informative parameters, line shape amplitudes, widths, and locations, for each constituent line shape along with predictive distributions for the the estimated resonant Raman signal contained in the CARS measurement spectrum, the error corrected CARS measurements, and the CARS measurement spectrum as well as extending currently existing methodology for modelling experimental artefacts present in CARS measurements.
Additionally, a line narrowing algorithm requiring little to no \textit{a priori} information on the underlying line shapes readily applicable to various spectral measurements is developed and is successfully used to obtain informative priors purely from the measurement data for the Bayesian inference model.
The applicability of the methods is demonstrated with experimental CARS spectra of sucrose, fructose, glucose, and adenosine phosphate.

\begin{acknowledgement}

The authors thank Prof.\ Heikki Haario for useful discussions and  Michiel M\"uller and Hilde Rinia for providing the experimental data.
This work has been funded by the Academy of Finland (project numbers 312122, 326341 and 327734).
MTM also thanks the Australian Research Council Centre of Excellence for Mathematical and Statistical Frontiers (project number CE140100049).

\end{acknowledgement}

\begin{suppinfo}
Model parameter priors and obtained posterior distributions of the parameters for each case are available online.
\end{suppinfo}


\providecommand{\latin}[1]{#1}
\makeatletter
\providecommand{\doi}
  {\begingroup\let\do\@makeother\dospecials
  \catcode`\{=1 \catcode`\}=2 \doi@aux}
\providecommand{\doi@aux}[1]{\endgroup\texttt{#1}}
\makeatother
\providecommand*\mcitethebibliography{\thebibliography}
\csname @ifundefined\endcsname{endmcitethebibliography}
  {\let\endmcitethebibliography\endthebibliography}{}

\end{document}